\documentstyle[seceq,wrapft,epsf]{ptptex}
\newcommand{\Psfig}[2]{\epsfysize=#1\epsfbox{#2}}

\title {
Fragment Distribution\\
in Coexistence Phase of Supernova Matter 
}

\author{Chikako {\sc Ishizuka}$^1$\footnote{E-mail:chikako@nucl.sci.hokudai.ac.jp}, 
\underline{Akira {\sc Ohnishi}}$^1$ and Kohsuke {\sc Sumiyoshi}$^2$
}
\inst{
$^1$	Division of Physics, Graduate School of Science\\
	Hokkaido University, Sapporo 060-0810, Japan
\\
$^2$ Numazu College of Technology, Numazu 410-8501, Japan
}
\recdate{
}

\begin{document}
\maketitle
\begin{abstract}
 We propose a new nucleosynthesis process 
 in which various nuclei are formed through the liquid-gas phase transition
 of supernova matter as a preprocess of the standard r-process,
 and study its possibility of realization qualitatively.
 In the relativistic mean field and statistical model calculations,
 we find that this process may take place
 and help the later r-process to proceed.
\end{abstract} 
\thispagestyle{empty}

\section{Introduction}
It is generally believed that there exist several phases of nuclear matter.
Among the phase transitions between them, the nuclear liquid-gas phase
transition has been extensively studied in these three decades.
It takes place at relatively cold ($T_c = (5-8)$ MeV)
and thin ($\rho_B \sim \rho_0/3$) nuclear matter, 
and it causes multifragmetation of expanding nuclear matter
in heavy-ion collisions~\cite{HeavyIon}.
In the universe, 
the temperature and density region of this transition
would be probed during supernova explosion.
Since supernova explosion can be regarded as expansion of infinite
matter composed of nucleons and leptons,
its fragmentation through the phase transition
may produce various nuclei including heavy nuclei
which are believed to be synthesized in r-process or rp-process.
Then the phase transition during supernova explosion may be helpful
to solve some problems in these standard processes.

 In this work, we propose a new kind of nuclear synthesis process through the 
 liquid-gas phase transition (LG process) as a preprocess of 
 the usual r-process.
 In the early stage of explosion, since the density and the temperature
 are high enough to keep statistical equilibrium and to trap neutrinos,
 entropy per baryon $S/B$ and lepton to baryon ratio $Y_L$ are to be
 conserved. 
 This condition is expected to be satisfied within the neutrino sphere
 ($\rho_B=(10^{-4}-1)\rho_0$).
 At these densities,
 supernova matter expands almost adiabatically and cools down.
 If the critical temperature $T_c$ of the liquid-gas phase transition
 is higher than the freeze-out temperature,
 this matter will fragment and form various nuclei in a critical manner
 through the phase transition.
 As the baryon density decreases,
 charged particle reactions are hindered
 then the chemical equilibirum ceases to be kept, 
 namely the system freezes out at this point.
 The statistical distribution of fragments at freeze out
 will give the initial condition of the later processes such as the r-process.

 Although the importance of the liquid-gas phase transition
 in supernova matter was already noticed and extensively studied
 before~\cite{Lattimer}, 
 the main interest in these works was limited
 to the modification of the equation of state (EoS),
 then the nuclear distribution as an initial condition
 of the r-process was not studied extensively.
 In addition, the mean field treatment assuming one
 dominant fragment configuration was applied rather than to 
 consider the statistical ensemble of various fragment configurations, 
 which is important at around the critical temperature
 for sub-saturation densities.

 In this study, we first investigate
 the liquid-gas coexistence region in the $(\rho_B, T)$ diagram
 and the ejection possibility of materials inside this coexistence region
 within the Relativistic Mean Field (RMF) theory.
 Next, by using a statistical model, 
 we evaluate the fragment formation at around the phase transition
 and in the coexistence region.
 Then we compare the fragment distributions at various $(\rho_B,T)$ points
 with the solar abundance.

\section{Relativistic Mean Field}
\begin{wrapfigure}[20]{r}{\halftext}
\centerline{\Psfig{6cm}{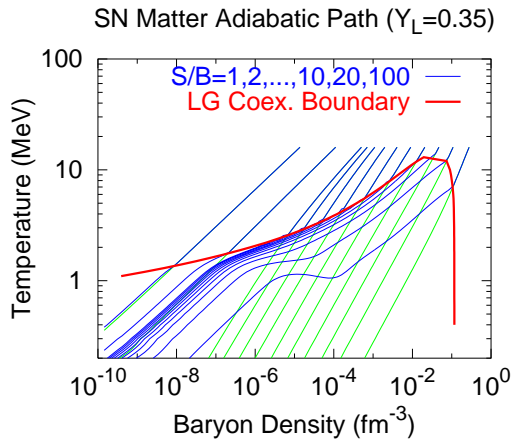}}
\caption{
 The boundary of the liquid-gas coexistence region  
 of supernova matter (thick solid line), 
 and the adiabatic paths with coexistence (thin solid lines)
 and without coexistence (uniform matter, dotted lines).
}
\label{Fig.1}
\end{wrapfigure}
 Here, we study the phase diagram of supernova matter
 by using RMF theory.
 For simplicity, we have made following three assumptions:
 (1) Supernova matter is composed of nucleons, electrons, electron-neutrinos
     and their antiparticles.
 (2) Neutrinos are trapped then the lepton to baryon ratio is conserved.
 (3) Two coexisting (liquid, gas) phases are of infinite size.
%
%
 Based on these assumptions, we have carried out the RMF calculation
 with the interaction parameter set TM1~\cite{TM1}, 
 which reproduces the binding energies of stable as well as unstable nuclei.
%
%
 In supernova matter, there are three conserved quantities
 (baryon and lepton numbers and charge), then 
 three different chemical potentials are required.
 In order to get thermal and chemical equilibrium with the chemical 
 potentials, we solve two phase balance under the Gibbs condition.

%
%
 In Fig.1, we show the boundary of the liquid-gas coexistence region  
 of supernova matter and the adiabatic paths at $Y_L = 0.35$.
 First, one finds that the coexistence region extends 
 up to $T_c \sim 15$ MeV at maximum,
 and remains up to $T_c \sim 1$ MeV even at very low densities.
 These critical temperatures are much higher than those in asymmetric nuclear
 matter where $Y_p$ is fixed, 
 and comparable to those in the symmetric nuclear matter.
 This large $T_c$ is due to leptons:
 Since $Y_L$ is kept constant but $Y_p$ is free from constraint, 
 the supernova matter searches its minimum free energy state
 by changing $Y_p$~\cite{IOS2001-YKIS01b}.

 Second, all the adabatic paths seem to go across the boundary of 
 the coexistence region.
 A Hydrodynamical model calculation
 shows that materials with $S/B \geq 10$ are to be ejected to outer space.
 The EoS in this hydrodynamical calculation
 is given by using the TM1 interaction.
 Therefore,
 the present model calculation suggests that
 high entropy supernova matter would be ejected after going through
 the coexistence phase, 
 provided that the freeze-out temperature of supernova matter is 
 as low as $T = (1-2)$ MeV
 at densities $\rho_B = (10^{-10}-10^{-5}) \hbox{fm}^{-3}$
 and neutrinos are still trapped at these densities.
 Although the first requirement may be satisfied,
 the densities of neutrino trapping in the time scale of supernova explosion
 are estimated to be $\rho_B \geq 10^{-5} \hbox{fm}^{-3}$.
 From this point of view, it would be interesting to find 
 that the adiabatic path with $S/B \sim 10$ crosses this boundary
 at around $(\rho_B, T) = (10^{-5}\,\hbox{fm}^{-3}, 1.5 \hbox{MeV})$,
 which satisfies both of the requirements.

 Above observations tell us that it would be marginal but still possible
 for supernova matter experienced the liquid-gas phase transition
 to be ejected to the outer space.
 However, in the present model, the treatment of fragments is not realistic,
 since we have assumed that the coexisting two phases are infinite.
 Nuclear matter in the liquid phase is more symmetric than in the gas phase
 then both of these are charged, then we cannot neglect the Coulomb energy.
 Therefore, in the next section, we evaluate the fragment yield 
 in a fragment-based statistical model.

\section{Statistical Model of Fragments}
 In order to deal with the distributions of nuclei having finite size,
 we choose a statistical model used in heavy-ion collision study,
 which is almost the same as the 
 Nuclear Statistical Equilibrium (NSE) used in the field of astrophysics.
 In the calculation, we solve
 the statistical equilibrium among nucleons, fragments, and leptons
 with fragment-based grand canonical ensemble.
 The distribution function is given as follows.  
\begin{eqnarray}
\rho_f(A,Z)
	&=& g(T)\int\frac{d^3p}{(2\pi\hbar)^3}
		\exp\left(-\left(E_f-\mu_f\right)/T\right)
\ , \\
E_f
	&=& \frac{p^2}{2M_f}-B_f(A,Z)+V_c(A,Z)
\ ,\quad
\mu_f
	= A\mu_B + Z\mu_c\ ,
\end{eqnarray}
 where $E_f$ and $B_f$ denote the energy and the binding energy
 of each nucleus, and $V_c$ is the Coulomb potential energy between fragments.
 The spin degeneracy factor $g(T)$ contains the contributions
 from ground state as well as from excited states estimated
 in the Bethe formula.

 In heavy-ion collision studies,
 the experimental values are used for the binding energies
 and only observed nuclei are considered as fragment species.
 In supernova matter, however, several modificiations are required.
 First, since the electron density is much higher in supernova matter,
 the Coulomb energy is screened and reduced.
 We here assume that electrons are uniformely distributed in a sphere
 with radius $R_e$ which is determined to cancel the charge of the nucleus
 at a given electron density.
 Then the Coulomb energy correction is given as
\begin{equation}
\Delta V_c(A,Z)
	= - a_c\frac{Z^2}{A}
	\left(\frac{3}{2}\eta - \frac{1}{2}\eta^3\right)
\ ,\quad
\eta
	= R_A/R_e \propto \rho_e^{1/3}
\ .
\end{equation}
 Because of the functional form $\rho_e^{1/3}$, 
 this correction is meaningfully large even at $\rho_B = 10^{-6} \rho_0$.
 For example, when the electron-baryon ratio is $Y_e = 0.3$ 
 the corrections amount to
 90 \% of the total Coulomb energy and 10 MeV
 at $\rho_B = \rho_0$ and $\rho_B = 10^{-6}\rho_0$, respectively.
 Therefore we cannot neglect this Coulomb energy correction by electrons
 (i.e. screening by electron) even at low densities.
 In addition, because of the finite gas nucleon density,
 nuclei outside of the dripline might be stabilized.
 Then we have adopted the mass table of Myers and Swiatecki~\cite{MS1994}
 which contains 9000 kinds of nuclei.

 By using the extended statistical model which uses the mass table
 of Ref.~\citen{MS1994} and includes the Coulomb energy correction,
 first we have estimated the critical temperatures at several densities,
 as shown in Table I, from the kink in the caloric curve
 and the power law fit of the fragment mass distribution.
 If the density is not very small,
 the freeze-out temperature is expected to be
 around or just below the critical temperature.
 Below the critical temperature, fragments are formed abundantly
 and nuclear number density becomes much smaller,
 then the mean free path for each nucleus becomes much longer.
 This rapid fragment formation makes the typical interaction intervals longer,
 and is expected to help the system to freeze-out.

\begin{figure}
\begin{minipage}{0.5\linewidth}
\centerline{\Psfig{5.5cm}{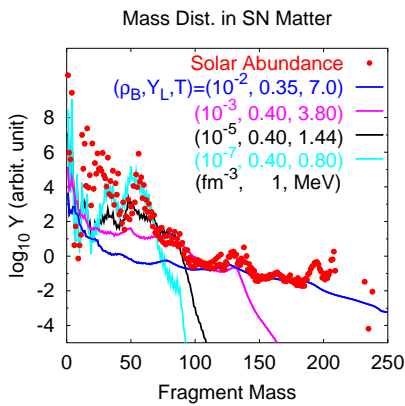}}
\caption{Mass distribution in supernova matter.
}
\label{Fig.2}
\end{minipage}
\begin{minipage}{0.5\linewidth}
\centerline{\Psfig{5.5cm}{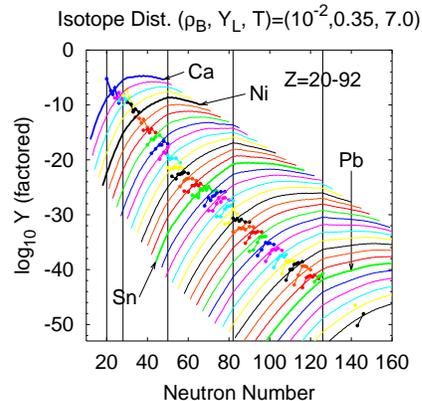}}
\caption{Isotope distribution in supernova matter.
}
\label{Fig.3}
\end{minipage}
\end{figure}

 In Fig. 2, we show the fragment mass distributions at temperatures
 just below $T_c$ for several baryon densities.
 When the freeze-out density is as low as $\rho_B = 10^{-7} \hbox{fm}^{-3}$, 
 nucleons form nuclei around the iron peak
 and a small number of the first peak nuclei of the r-process at equilibrium.
 As the freeze-out density increases,
 heavier nuclei are formed more abundantly.
 For example, nuclei around the first, second and the third peak of the
 r-process can be formed at freeze-out densities of
 $10^{-5}, 10^{-3}$, and $10^{-2} \hbox{fm}^{-3}$, respectively.

\begin{table}
\caption{
Temperatures and typical formed nuclei shown in Fig. 2
for several baryon densities.
The critical temperatures are written in the bracket.
}
\begin{center}
\begin{tabular}{c|c|c}
\hline
\hline
Baryon Density (fm$^{-3}$) & Temperature ($T_c$) (MeV) & Formed Nuclei\\ 
\hline
 $10^{-2}$  &  7.0~~( 7.2) & 3rd Peak 	   \\
 $10^{-3}$  &  3.8~~( 4.0) & 2nd Peak 	   \\
 $10^{-5}$  & 1.44~~(1.48) & 1st Peak	   \\
 $10^{-7}$  & 0.80~~(0.84) & Fe,Ni + 1st Peak \\ \hline
\end{tabular}
\end{center}
\end{table}
 It is interesting to find that, 
 at high freeze-out densities,
 the r-process peak structure is not clear and shifted to lower mass direction.
 These broadness and shifts can be well understood in the isotope distribution.
 In Fig. 3, we show the calculated even $Z$ isotope distribution from Ca to U
 in the present statistical model at the highest freeze-out density,
 $\rho_B = 10^{-2} \hbox{fm}^{-3}$, in comparison with the solar abundance.
 One overall normalization factor is multiplied to get a good global fit.
 We find that calculated values fit the observed (long-lived) isotope
 distribution very well from Ca to U.
 We also find that there are a huge number of very neutron rich nuclei
 appears in the calculated results.
 In the r-process, neutron absorption process stalls at the neutron
 magic numbers, but in high density supernova matter, 
 nuclei beyond the drip line is statistically stabilized
 because of the large neutron chemical potential.
 Similar trends can be seen in the isotope distribution at lower 
 freeze-out densities, although the explained range of $Z$ is limited
 to $Z <$ 56 (Ba), 40 (Zr), and 32 (Ge) for 
 $\rho_B = 10^{-3}, 10^{-5},$ and $10^{-7} \hbox{fm}^{-3}$.
 After freeze-out,
 these very neutron rich nuclei will provide a lot of
 neutrons by neutron emission.

\section{Summary and Discussion}
 In this paper, we have proposed the LG process
 which may play an important role for the understanding of 
 mass and isotope distribution in the unverse.
 In order to evaluate its possibility qualitatively,
 we have demonstrated that the coexistence phase extends to low densities
 keeping the critical temperatures to be greater than 1 MeV, 
 then materials in the coexistence region
 can be ejected to the outer space at supernova within a two phase model of RMF.
 Next, we have performed a statistical model calculation which includes
 the mass table extension and the Coulomb energy correction
 to adapt to supernova matter.
 The calculated results show that nuclei above the iron peak would be formed
 in the coexistence region at relatively high freeze-out densities,
 and at these densities, observed isotope distribution 
 are explained well.

Works for a more quantitative discussion are in progress.
%
%
%



\begin{thebibliography}{9}
\bibitem{HeavyIon}
  J.~Pochadzalla et~al. (ALADIN Collab.),
	Phys. Rev. Lett. {\bf 75} (1995), 1040;
  W. Reisdorf et al. (FOPI Collab.),
	Nucl. Phys. {\bf A612} (1997), 493;
  R. Nebauer et al. (INDRA Collab.), 
	Nucl. Phys. {\bf A658} (1999), 67;
  M. D'Agostino et al.,
	Phys. Lett. {\bf B473} (2000), 219.
\bibitem{Lattimer}
  D.Q. Lamb, J.M. Lattimer, C.J. Petheck and D.G. Ravenhall
	Nucl. Phys. {\bf A360} (1981), 459.
\bibitem{TM1}
  Y. Sugahara, H. Toki,
	Nucl. Phys. {\bf A579} (1994), 557. 
\bibitem{IOS2001-YKIS01b}
  C. Ishizuka, A. Ohnishi, K. Sumiyoshi,
	in this proceedings.
\bibitem{MS1994}
  W.D. Myers and W.J. Swiatecki,
  	LBL-36803 (1994); Ann. of Phys. {\bf 204} (1990), 401.
\end{thebibliography}
\end{document}